# Native point defects in CaS: A focus on doping limit for persistent luminescence

Bolong Huang

*Department of Physics and Materials Science, City University of Hong Kong, Kowloon, Hong Kong SAR, China*

**Abstract**

We studied native point defects in CaS by DFT+ Hubbard U method. The effect of the localization of the d orbitals of Ca pseudopotential has been included. The Hubbard U corrected d-orbital for Ca sites are playing a role assisting the charge transfer from p orbitals of S site to the Ca site, giving both localized electron and hole states within the band gap, with an energy interval of 1.2~1.4 eV. This corresponds to the localized excitonic levels below the conduction band edge for the optical absorption. The electronic properties and formation energies of native point defects have been discussed. We found the neutral S vacancy has the lowest energy of 0.62 eV under Ca-rich limit. The Schottky defect pair defect is another dominant defect with cost of 1.51 eV per defect site from S-rich to Ca-rich chemical potential limits. The defect formation energy further summarized that the doubly positive S vacancy and doubly negative Ca vacancy are both the most stable donor-type and acceptor-type defects respectively coexisting in CaS. We also summarized a narrow doping limit energy which has been determined as 1.33 eV constantly in CaS independent to different chemical potential limits. But such doping allowed range entirely shifts from valence band vicinity toward the conduction band edge from S-rich to Ca-rich limits. Under S-rich limit, the dopable range for $E_F$ shifting corresponds to $Eu^{2+}$ doping experiments, the luminescence wavelength limit is predicted to be 659 nm (1.88 eV) remarkably close to the reported results 650 nm. This gives a solid theoretical reference for the lanthanide ions doping experiments in CaS. We conclude that the defect levels combined with formation energy is an accessible way to suggest the doping energy range and an assistant to explain the activation stage of photostimulated luminescence (PSL) mechanism in lanthanide ions doped crystal materials.

**Introduction**

The red long-persistent luminescence technique has been recently attracted tremendous interests in biological, chemical and physical applications[1-4], which is related to the materials of storage phosphor. The storage phosphor like calcium sulfide (CaS) is a wide-gap insulator and ionic compounds. As reviewed by Smet et al, the synthesis of CaS as a good phosphor can be found as early as 1700 by Friedrich Hoffmann[5]. However, even to date, there still lacks of comprehensive understand on the intrinsic electronic and optical properties of crystal CaS in related to its native point defects. As we believe, the native point defects play significant role on affecting its luminescence properties of ionic compounds[6].

Recently, Capobianco and co-workers reported the persistent and red photostimulated luminescence (PSL) based on both CaS:$Eu^{2+}$, $Dy^{3+}$ upon UV and 980 nm photo-irradiations, respectively[4]. Such persistent luminescence can be explained by the carriers (electrons/holes) at the shallow trap levels conduction band minimum (CBM) stimulated into the conduction band by the photo irradiation, and recombined at the acceptor levels given by cation dopants like $Eu^{2+}$ and $Dy^{3+}$. This requires the intrinsic trap levels are around 1.2~1.5 eV lower than the CBM to be efficiently activated by 980 nm irradiation for carriers (say electrons here). Capobianco et al also deduced the similar intrinsic trap level (~1.3 eV) through their luminescence measurements[4]. X-ray powder diffraction (XRPD) results confirm the simple cubic symmetry (*Fm-3m*) for what the synthesized nanoparticle CaS:$Eu^{2+}$ possesses[4]. Therefore, such advanced optical materials applications urge us to have a preliminary understand on its native defect states of crystal CaS.

As is well known, many ionic compounds like alkli halide as NaCl and LiF or oxides like CaO, and MgO [7] have a variety of color emissions by the F centers induced by anion vacancies. Such F center contains the natively trapped electrons within the band gap and their electronic trapping levels determine the optical emission properties. For both successful persistent and PSL, it is required us to carry out a first-principle study to answer whether the native F center of CaS meets the activation energy level requirement of 1.2~1.3 eV.

Before the study on the electronic and optical properties influenced by the native defects states, it is necessary to obtain a detail knowledge of the electronic structure of CaS, in particular the nature of the unoccupied density of states near the conduction band edge or minimum (CBM), as this arise much attention on the consideration of the 3d orbitals of Ca into valence electron wavefunctions into ab-initio calculations. Whether the 3d orbitals of Ca are to be treated as fully empty or partially filled (e.g. $3d^{\delta}$ with $0<\delta<1$) is also question to be carefully taken into account.

The point defects are of great importance in controlling the creation and annihilation of free charge carriers in the solid materials. CaS is a typical insulating materials for photostimulated luminescence application that dominated by intrinsic point defects. The systematic study on the intrinsic defect levels that control the electronic and optical properties is still no clear and even unknown. This motivated us a strong intention in predicting their properties and explaining the experimental observations.

The three main interesting points for defect in the solids are: (i) the defect formation energy $H(\mu, E_F)$ is a function of the Fermi energy ($E_F$) and chemical potential μ of the solid components. The formation energy decides the defect or doping concentration reachable in the solids at a given temperature as well as the $E_F$ in equilibrium. (ii) The transition energy, is the energy E(q, q') required to ionize the defect or dopant center from charge state q to q'. It is also called donor or acceptor transition energy which presents the ability of the defect or dopants to produce the free charge carriers at a given temperature. (iii) The charge localization of the defect or dopants that show localization of carriers. To our knowledge, the acceptors in solid materials often introduce hole states localized on single anion site but wrongly described by LDA. Clark and Robertson et al have correctly confirmed the localization by sX-LDA[8-10].

In the past few years, many advance first-principles methods like DFT, DFT+U, hybrid functionals, GW methods have been well performed on many solids. However, for DFT+U, based on self-consistently determined Hubbard U on cation's d-orbital and anion's p-orbital, the self-interaction error (SIE) can be reduced into an acceptably low level that recovers behaviors of both localized electrons and holes states. Note that, to accurately describe the electron/hole interactions that bounds the F-center trapping level in the solids, an appropriate magnitude of the hole localization on anion sites is required. This helps us to obtain accurate F-center excitation energies to meet the photo-irradiation in persistent and PSL studies[4, 5].

**Calculation setting**

We chose the cubic lattice with the same symmetry of rocksalt (NaCl) for modeling the crystal CaS. The GGA+U with rotational invariant scheme [11]was used from the CASTEP code[12]. The norm-conserving pseudopotentials of Ca and S are generated by OPIUM code in the Kleinman-Bylander projector form[13] and already employed non-linear partial core correction[14] to reduce the atomic core-valence electron densities overlap. We treated the (3*s*, 3*p*, 3*d*, 4*s*) states as valence states of Ca. The RRKJ method is chosen as optimization of pseudopotentials[15]. As we know the norm-conserving pseudopotentials can reflect all-electron behavior for outter shell valence electrons with **|S-matrix|**=1 compared to ultrasoft pseudopotentials[16, 17]. Since both the Ca and S are relatively light elements, the spin-orbit coupling effect has not been considered all over the calculations. The PBE functional was chosen for PBE+U calculations with a kinetic cutoff energy of 850eV, which expands the valence electrons states in a plane-wave basis set. The ensemble DFT (EDFT) method of Marzari et al[18] is used for reducing the charge-sloshing effect during the electronic minimization of solving Kohn-Sham equation. Reciprocal space integration was performed by *k*-point grids of $10 \times 10 \times 10$ *k* points in the Brillouin zone of CaS, respectively. This converges the total energy to under $5.0 \times 10^{-7}$eV per atom. The Hellmann-Feynman force on each atom was converged to lower than 0.01eV/Å.

We follow the Anisimov type DFT+U method[11] and the self-consistently determined Hubbard U parameter for Ca 3d orbital by our new linear response method. To stabilize the hole states lying in the S 3p orbitals, we also apply a self-consistent determined Hubbard U potential (method used above) to the S 3p states following Lany[10, 19], Morgan et al[20], and Keating et al[21]. Accordingly, both the d- and p- orbital electrons of the cations and anions should be considered when using DFT+U[22]. For defects, we use a 2x2x2 CaS supercell containing 64 atoms. We select the (¼, ¼, ¼) special k-point[23] in the simple cubic 2x2x2 supercell. The geometry optimization used the Broyden-Fletcher-Goldfarb-Shannon (BFGS) algorithm through all bulk and defect supercell calculations.

Another key setting is the pseudopotential for Ca. We generated the norm-conserving pseudopotential orbitals with $3s^2 3p^6 4s^2$ for Ca by OPIUM code. The $3d^0$ configuration is also considered into the generation. This does not change the valence distribution but will increase the accuracy of the valence electron interaction of Ca. In real DFT calculation, the $3d^0$ and $4s^2$ of Ca will have a certain overlapping with occupancy of 0.4~0.5 *e* through Mulliken analysis. This renders us to consider an on-site Hubbard U energy reserved for this fractionally occupied empty 3d orbital for Ca. Such consideration of $3d^\delta$ ($0<\delta<1$) is also verified in the electronic structure calculations and experiments of $CaB_6$, which is a new semiconductor for spin electronics[24-26]. In that the structure has 3d of Ca hybridized with 2p orbitals of the anions affecting the band gap size of Brillouin zone[27]. The total plane wave basis set for Ca, we use the 850 eV that was suggested by Clark et al with similar norm-conserving generations[28]. With our self-consistently determination process[29], the on-site Hubbard U parameters for 3d of Ca and 2p of S are 2.51 eV and 5.14 eV, respectively.

For the calculation of defect formation energy in different charge states, the overall supercell size was kept fixed based on the relaxed neutral bulk unit cell. The defect formation energy ($H_q$) at the charge state $q$ as a function of the Fermi energy ($E_F$) and the chemical potential $\Delta\mu$ of element $\alpha$ is given by

$$H_q(E_F,\mu) = [E_q - E_H] + q(E_V + \Delta E_F) + \sum_\alpha n_\alpha(\mu_\alpha^0 + \Delta\mu_\alpha), \tag{1}$$

where $E_q$ and $E_H$ are the total energy of a defect cell and a perfect cell, respectively, calculated of charge $q$, $\Delta E_F$ is the Fermi energy with respect to the valence band maximum, $n_\alpha$ is the number of atoms of element $\alpha$, and $\mu_\alpha^0$ is reference chemical potential, following Lany and Zunger[30].

**Results and discussion**
**Bulk CaS**

CaS has a rock-salt structure with symmetry of Fm-3m with lattice parameter of about 5.70 Å. The relaxed structure of ground state by DFT+U has a lattice parameter of 5.84 Å with small error of +2%. From Figure 1, the electronic band structures calculations based on the optimized cell got a 5.13 eV indirect band gap from Γ to X transition, close to the results reported by Capobianco et al[4]. The direct transition gap of Γ→Γ is 5.55 eV close to the result of 5.57 eV reported by Poncé et al using GW method[31]. Another direct transition gap of X→X is 6.14 eV with 1.0 eV higher than the results from Poncé et al[31]. The valence band width we got is 3.25 eV and the conduction band width is 3.69 eV. The valence band (from -3.3 to 0 eV in Figure 1) is contributed by 3p orbitals of S, while the full-filled 3p level of Ca is down to the -8 eV relative to the highest occupied level (0 eV). The conduction band is mainly contributed by 3d and 4s orbitals of Ca, and lowest unoccupied level is dominant by 3d level of Ca. These are remarkably consistent with the results calculated by both Tran-Blaha 09 (TB09) and many-body GW method from Waroquier et al[32]. Moreover, we see that with both applied Hubbard U parameters on the 3d of Ca and 3p of S, the electronic structure calculation not only substantially improves the band gap, but also the nearly consistent valence and conduction band widths, respectively. This is also can be seen from the other work previously[22].

For the bulk formation enthalpy of crystal CaS, $\mu_{Ca} + \mu_S = \Delta H_f(CaS)$, the experimental formation enthalpy of CaS is reported as -5.00 eV at T=298K (-482.4 kJ/mol)[33]. Our calculations give -5.01 eV by GGA+U ($U_d$=2.51, $U_p$=5.14) at T=0K (ground state) which agrees well with the experimental value at T=298K.

**S vacancy ($V_S$)**

The $V_S$ could be seen as F center defect in CaS. The $V_S$ with charge states of 0, +1, and +2 has been investigated. The neutral $V_S^0$ left two electrons lying at nearby Ca sites with anti-ferromagnetic (AFM) behavior for the electronic states all over the Brillouin zone, shown in Figure 2 (a). The two localized states showing spin-up and down with 1.72 eV lower than the conduction band minimum (CBM). These two states are indeed provided by the two electrons fallen into the overlapped 3d-4s orbitals of neighboring Ca, which are originally left by S vacancy ($V_S$). Their energy levels are degenerated and hence giving an s-like wavefunctions. These overlapped 3d-4s orbitals of Ca in fact form the trapping levels that carry the excess electrons at the defect sites. This is very different to the other ionic compounds like alkli halide (NaCl, LiF etc.) which only consists s-p orbitals[34].

The $V_S^+$ also gives two localized states within the band gap but with one empty, the orbitals of the states are also given by the 3d-4s orbitals. As there is only one electron at the $V_S^+$ site, these 3d-4s trapping levels split due to the different electron occupancies, shown in Figure 2 (a). The filled trap level is 1.23 eV lower than the empty trap level, while the empty trap level is 1.16 eV lower than the CBM. Different from the TDOS of $V_S^0$, the singly positive $V_S^+$ presents a ferromagnetic (FM) behavior due to the absence of energy degeneration of the 3d-4s trap levels for the localized electrons of $V_S$ site. On the other hand, the empty state is actually the trap levels of the localized hole states in CaS with $V_S$, and the hole trap levels is actually 1.23 eV higher than the electron trap level at the ground states, and *vice versa* for the excitation states like 980 nm photo-irradiation experimentally (means the electron level is 1.23 eV higher than the hole level). Gao et al has studied the similar $CaB_6$ which is another Ca-based ionic compounds, their more accurate electron energy loss spectroscopy (EELS) experiments suggested that the Ca with d orbitals actively plays as ionic hole that transfer the charge from the s, p and d states[27]. This gives an evidence to prove our explanations of the split of the two trap levels.

The $V_S^{2+}$ turns back to the AFM state with two aligned spin-up and down empty states within the band gap. Both of them also return to the degenerated energy levels for the 3d-4s trap level. Those empty trap levels are sitting 1.67 eV below the CBM. Another evident change for the electronic states of the $V_S$ in CaS is the band edges from the TDOS in Figure 2 (a). The valence band edge is almost unchanged within those three different charge states (0, +1, and +2), while the conduction band edges have obvious variations near the CBM. Thus, we deduce that the localized electrons left by $V_S$ are only interacting with the 3d and 4s orbitals of Ca, and have no interactions and effects on the 3p orbitals of nearby S sites. These electrons trapped at the $V_S$ sites are in fact from the neighboring Ca ions. In all, the lattice and electronic behaviors of $V_S$ in CaS match the F center in ionic compounds, shown in Figure 2 (b) by schematic diagrams. Owing to the structural feature, the nearest neighbors of $V_S$ are all Ca site, we can see from Figure 2 (c) that, the $V_S^0$ has no lattice distortion effects, but the $V_S^+$ makes an evident distortion that pushes the six Ca sites outward while attracts the nearest S sites little inward, The $V_S^{2+}$ shows the distortion more vastly. The localized electron orbitals all have s-like wavefunction feature, while the localized hole state shown some extent of hybridized d-orbitals.

From the angle of defect formation energies, Figure 2 (d) shows the $V_S$ in CaS is a positive-$U_{eff}$ defect ($U_{eff}$ = +0.75 eV) , which is the energy cost of thermally ionizing charge states of $V_S$ cannot be compensated by the local lattice distortions. Under S-rich chemical potential limit, the $V_S^0$ cost 5.63 eV to form while it has an as low as 0.62 eV formation energy under Ca-rich limit. For the $F^+$ center defect, $V_S^+$, has energies of 1.76 eV and -3.25 eV under S-rich and Ca-rich limits, respectively. The $V_S$ has transition levels of (+2/+1) and (+1/0) states are 3.12 eV and 3.87 eV, respectively, which are 2.01 eV and 1.26 eV below the CBM. By the aspect of formation energy, the (+1/0) state in fact matches the transition level requirement of the persistent luminescence predicted by Capobianco et al[4].

**S interstitial ($I_S$)**
The $I_S$ in the CaS could be a strong acceptor center that gives the localized hole states can trap, but it could be an localized electron-hole trapping complex depends on the local bonding feature. For instance, the O interstitials ($I_O$) in metal oxides may form peroxides with O-O homo-polar bonds in the lattice, whose π-electrons localized along the O-O bond at the $I_O$ sites, show in previous work on metal oxides[22]. The neutral $I_S$ in CaS with S-S bond induces four localized levels with spin-degenerated shown in Figure3 (a). One of the localized electronic trap levels is at about 0.2 eV higher than the valence band maximum (VBM). The other two levels are -4.3 eV and -5.0 eV, below the VBM. They all have π-like electronic orbital features. The fourth localize 0.4 eV below the CBM presenting as a π* orbital for the anti-bonding state of the S-S bond by $I_S$.

For the singly negative $I_S$ in CaS ($I_S^-$), in the band gap shown in the Figure 3 (a), there are four gap states for trapping the electrons and holes. Two lined as spin-up and down with about 0.7 eV above the VBM, the second spin-up trap levels for the localized electron locates 3.1 eV higher than the VBM, while with 1.39 eV below the second spin-down trap states for the localized hole.

Figure 3 (b) shows that the localized electron and hole orbitals are all along the S-S bond direction induced by relaxed $I_S^-$ sites. The localization of the orbitals are more evident than the neutral $I_S$ site. While for the doubly negative $I_S^{2-}$, the relaxed structure is totally different, it sits in the center of a local cubic motif of CaS. The local tetrahedral bonding is pushing the neighboring S ions out of the local cubic motifs at the $I_S^{2-}$ sites shown in Figure 3 (b).

The $I_S$ in CaS is a negative-$U_{eff}$ defect center at the transition state of (-2/0) with -1.26 eV for $U_{eff}$. This means the chemical reaction process for $2I_S^- \rightarrow I_S^0 + I_S^{2-}$ is an exothermal process. The electron spin resonance (ESR) measurement would find a very low density of $I_S^-$ defect, since there only $I_S^0$ and $I_S^{2-}$ stably existing in the lattice. The energy cost of these defects formation can be compensated by the local lattice distortions.

We see from Figure 3 (c), under S-rich potential limit, the $I_S^0$ has energy of 2.06 eV to form in the CaS lattice, while 6.23 eV for $I_S^-$ under S-rich limit. The $I_S^-$ has similar trap levels in TDOS shown in Figure 3 (a), while the formation energy (Figure 3 (c)) shows a very low possibility for $I_S^-$ to singly form in the lattice. This is consistent to our deduced results from the negative-$U_{eff}$ for (-2/0) state.

**Ca vacancy ($V_{Ca}$)**
The $V_{Ca}$ in CaS left two localized holes in the lattice staying at neighboring S sites. We see in Figure 4 (a) that there are two localized single particle levels close in the band gap with 2.2 eV below the CBM and 2.83 eV higher than VBM, which are deep levels. The current calculations are able to reflect the two holes are trapped at the two nearby S sites, shown in Figure 4 (b).

For the singly negative Ca vacancy ($V_{Ca}^-$), there is only one localized hole left within the band gap, with the trap level of 2.2 eV unchanged. For the doubly negative $V_{Ca}^{2-}$ site in CaS, there is no trap levels in the band gap as no holes left in the lattice. We also find from the partial density of states (PDOS) that the local trap states are induced by the local p-orbitals. These holes are localized on p-orbitals and lined up along the z-axis of S sites in CaS, shown in Figure 4 (b).

From Figure 4 (c), the $V_{Ca}$ is a positive-$U_{eff}$ defect, which is about 0.14 eV. The (-/0) shows an acceptor feature with 0.57 eV higher than the VBM. The difference of this energy level to the single particle trap level arise because the (-/0) is the thermal ionization energy for the charge state varies from -1 to 0, while the single particle level is the electron/holes occupation energy level in the band gap. The $V_{Ca}^0$ in CaS takes 2.74 eV and 7.76 eV to form under the S-rich and Ca-rich potential limits, respectively.

**Ca interstitial ($I_{Ca}$)**

The excess Ca in CaS ($I_{Ca}$) carries donor-like electrons. The neighboring S will form strong hybridized bonds to the $I_{Ca}$. We see from Figure 5 (a), the donor-like level is 0.4 eV below the CBM for the $I_{Ca}^0$ with two electrons, while it shows 0.3 eV level below the CBM for singly positive $I_{Ca}^+$ with one electron occupancy. The doubly positive $I_{Ca}^{2+}$ does not shown any trap levels within the band gap. From Figure 5 (b), we see that the $I_{Ca}^0$ actually has two spin-up $d_{z^2}$-like electrons localized at the $I_{Ca}^0$. This shows the interstitial Ca with $3d^\delta 4s^{2-\delta}$ ($\delta>0$) in the CaS lattice actually donating its two electrons localized on d orbitals. For the $I_{Ca}^+$, the localized orbital is strongly hybridized between $d_{yz}$ and p orbitals (from S sites).

Figure 5 (c) shows the formation energies of the $I_{Ca}$ in 0, +1, and +2 charge states with different chemical potential limits. We see that the transition level of both (+2/+1) and (+1/0) states are in fact entering into the conduction band with about 0.4 eV. We also find that there is a positive-$U_{eff}$ of 0.21 eV. For the neutral state, the $I_{Ca}$ shows energetically unfavorable in CaS lattice, since the formation energy is 12.47 eV and 7.46 eV at the S-rich and Ca-rich limits, respectively. For the $I_{Ca}^+$, the formation energy turns down to the 6.90 eV and 1.88 eV at the S-rich and Ca-rich limits, while it is still unknown for experiments that actualizing the $Ca^+$ injection.

**Frenkel defects**

The Frenkel defect denotes one ion escape from the original site and trapped in the lattice, which is a complex of vacancy and interstitial of the same ion. The Ca Frenkel defect we say is c-Fr which means the cation related. The S Frenkel is given as a-Fr showing anion related. We firstly look at the a-Fr defect pair, since the anion related defect like vacancy ($V_S$) and interstitial ($I_S$) usually induced ground state s-like electron trap level and excited state d-like hole trap level.

For the neutral a-Fr defect, Figure 6 (a) shows the TDOS has ferromagnetic (FM) feature. The $V_S+I_S$ induce a p-π like band with about 0.6 eV higher than the VBM, induced by the excess 3p electrons of $I_S$. And those p-π electron localized along the S-S bond formed by the $I_S$ induced lattice distortions, shown in Figure 6 (b). The deep electron trap level is 2.38 eV below the CBM showing a deep donor-like state devoted by $V_S$ site with spin-up state. Another deep donor-like level with spin-down state is about 2.25 eV below the CBM is contributed by the $I_S$ site with p-π like orbital along the S-S bond direction, shown in Figure 6 (a) and (b). We also find that the two pairs with spin-up electron and spin-down hole states. The energies between the two pair states are 1.24 eV and 1.38 eV, respectively shown in Figure 6 (a). From the charge orbital analysis, the pair-states with 1.24 eV is contributed by $V_S$, the pair-states with 1.38 eV is given by $I_S$.

Formation energy study shows the a-Fr defect has 8.22 eV per pair defect which means 4.11 eV for each point defect. We show the formation energies of the a-Fr with charge states of -1, 0, and +1. We see the a-Fr is a negative-$U_{eff}$ defect center with energy of -1.48 eV for the exothermal process: $2(a\text{-}Fr)^0 \rightarrow (a\text{-}Fr)^+ + (a\text{-}Fr)^-$. The transition state of (-/+) stays at 3.61 eV in the band gap, shows the thermal ionization energy of a-$Fr^{(-/+)}$ defect pair is about 1.52 eV below the CBM. Therefore, the realistic a-Fr defect pair is existed as a-$Fr^{(-/+)}$ with formation energy of 3.74 eV per defect site, which is from the formation energy of the a-Fr$^0$ deduce by $|U_{eff}|/2$, learned from the Figure 6 (b).

**Schottky defects**

The Schottky defect in CaS is formed by the Ca and S mono-vacancies ($V_{Ca}+V_S$). In this type of defect, the two electrons left by $V_S$ at the Schottky defect site are neutralized by the two holes induced by $V_{Ca}$. Compared to the Frenkel defect pair, the Schottky defect pair has been found to be dominant in ionic compounds like alkaline-earth sulfides and other akali-halide by Pandey et al[35-37].

Figure 7 (a) shows the TDOS of the Schottky defect pair in CaS within different charge states, -1, 0 +1. The neutral state of such defect gives only the localized hole trap levels with 1.30 eV below the CBM, contributed by the d-orbitals of 5 nearby Ca sites. For the singly positive state, there are three localized holes within the band gap with two of them are spin-up and down in the same level. The energy intervals of the two trap level is 0.91 eV. The singly negative state has one localized electron and one localized hole in the band gap with the energy of 1.26 eV between these two states, and the localized hole state is 1.09 eV below the CBM.

We see from Figure 7 (b) that, the localized electron and hole state wavefunctions are similar to the case of $V_S^+$ discussed in previous content, which are all contributed by the d-orbitals of first neighboring Ca sites.

From the formation energy calculation, the Schottky defect pair has 3.01 eV per pair in neutral state, which means 1.51 eV per defect site. It is 2.23 eV lower than the a-Fr$^{(-/+)}$ defect pair discussed above. However, the Schottky defect pair has positive-$U_{eff}$ of 1.69 eV for the process: $2(STK)^0 \rightarrow (STK)^+ + (STK)^-$.

**Doping limit**

Figure 8 (a) and (b) summarized the native point defects in both S-rich and Ca-rich limits discussed above, which helps us to systematically analyze the doping in CaS using the concept of pinning energy rule. The pinning energy rule consists of n-type pinning energy and p-type pinning energy as elucidated by Robertson et al[38]. As introduced, the key limitation for doping in CaS is native defect compensation. This occurs because the Fermi level ($E_F$) moves to a specific band edge that causes the spontaneous formation of compensating defects due to the formation energy of the defect has fallen to zero at that $E_F$ position. We see that, for the neutral charge state, $V_S$ has the lowest formation energy in Ca-rich (S-poor) limit, as low as 0.62 eV. The neutral Schottky defect ($V_{Ca}+V_S$) has the second lowest formation energy of 1.51 eV per defect site, which presents another dominant defect existing in CaS.

We further discussed the doping limit energy determined by the native point defects of CaS. Figure 8 (c) further plots the formation energy of the most stable charge state of the most stable donor-type and acceptor type native defects for crystal CaS. It shows that the $V_S^{2+}$ and $V_{Ca}^{2-}$ are the most defects coexisting in CaS under both S-rich and Ca-rich limits. We firstly discussed the case of S-rich. If a donor is used to raise $E_F$ toward the conduction band edge (VBM, or $E_C$ in Figure), we must consider the formation energy of possible negatively charged compensating acceptors such as $V_{Ca}$ or $I_S$. Figure 8 (a) and (b) shows the $V_{Ca}$ is the more stable of these two defects, and the $E_F$ moves to 2.01 eV for $V_{Ca}^{2-}$ to form spontaneously. This energy is called the "n-type pinning energy" (or n-type limit) for donors[38, 39].

On the other hand, if an acceptor dopant is used to shift the $E_F$ to move lower energy toward to valence band edge $E_V$ (or VBM), we must look at the formation energy of compensating donor defects such as $V_S$ or $I_{Ca}$. For that, the $V_S$ has the lower energy of these two defects. Figure 8 (c) shows that in the case of S-rich limit, the $V_S$ has a negative formation energy if the $E_F$ drops below the 0.68 eV above the $E_V$. The $V_S$ (S vacancy) will spontaneously form if the $E_F$ moves toward $E_V$ and hinder the p-type doping by opposite charge compensation of $V_S$. The Fermi energy of 0.68 eV is the "p-type pinning energy" (or p-type limit)[38]. Therefore, in the case of S-rich, the enrgy range where we can shift $E_F$ by doping without spontaneously form opposite charge compensating defects is between these two limit energies, and yield as 1.33 eV.

The analysis is the same if applies to Ca-rich (or S-poor) conditions. Figure 8 (c) shows that both n- and p-type limit energies increase by 2.50 eV which is the half of the calculated formation enthalpy of crystal CaS discussed in earlier section, (the factor divided by is the charge of the defect, 2). The energy range for which any native compensating defect does not form spontaneously is still 1.33 eV, from 3.18 eV to 4.51 eV for $E_F$, determined by $V_S^{2+}$ and $V_{Ca}^{2-}$ respectively. Accordingly, the CaS can be dopable in both n-type and p-type but with narrow dopable range for $E_F$ variation, denotes the deep trap levels. The pinning-limit energies for doping in CaS is illustrated in a schematic band diagram as shown in Figure 8 (d). We set the 0 eV level is the vacuum level for the CaS, and the conduction band edge is decided by the calculated electron affinity energy (EA), which is 3.94 eV below the vacuum level. The n-type and p-type limit energy range has been shown as orange shaded area in Figure 8 (d). The calculated hope trapping level is shown as the exciton level with 1.24 eV below the conduction band edge ($E_V$). The energy between the n-type limit and exciton level is the 1.88 eV, which decides the upper limit of the photo-stimulated luminescence wavelength, 659 nm. This shows very close value to the experimental reported data by Capobianco et al[4].

**Summary**
We have given a detail discussion on the native point defects in the CaS crystal. We find that the d-orbtial of Ca sites in CaS has prominent effect on charge transfer showing a localized hole states for singly charged states of the native point defects of CaS. The localized electron and hole states have an energy intervals of about 1.2~1.4 eV, which is very consistent to the excitation energy by the 980 nm photo-irradiation for the persistent luminescence study of rare-earth doped CaS materials. The limits to doping in CaS have been calculated from the native point defects formation energies. The chemical trends of the native point defects and doping limits have been plotted with a band diagram together with the Fermi energies. As such, CaS has very narrow dopable range with 1.33 eV. There are advantages to us using the defect level study combined with formation energies to suggest doping energy limits and explain the photostiumulated luminescence in terms of native point defects.

**Figure 1.**

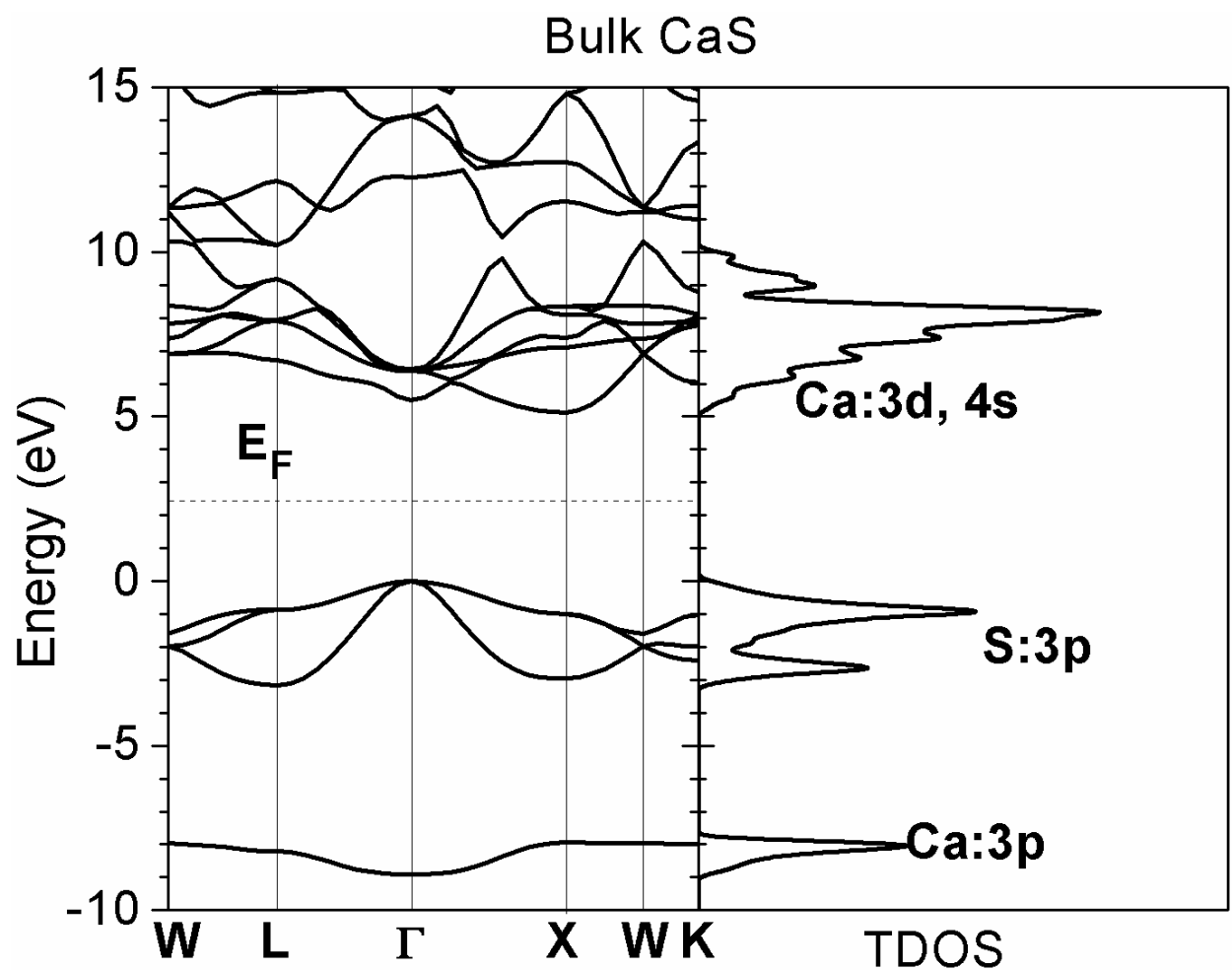

**Figure 1.** Band structure and TDOS of bulk CaS.

**Figure 2.**

**(a)**

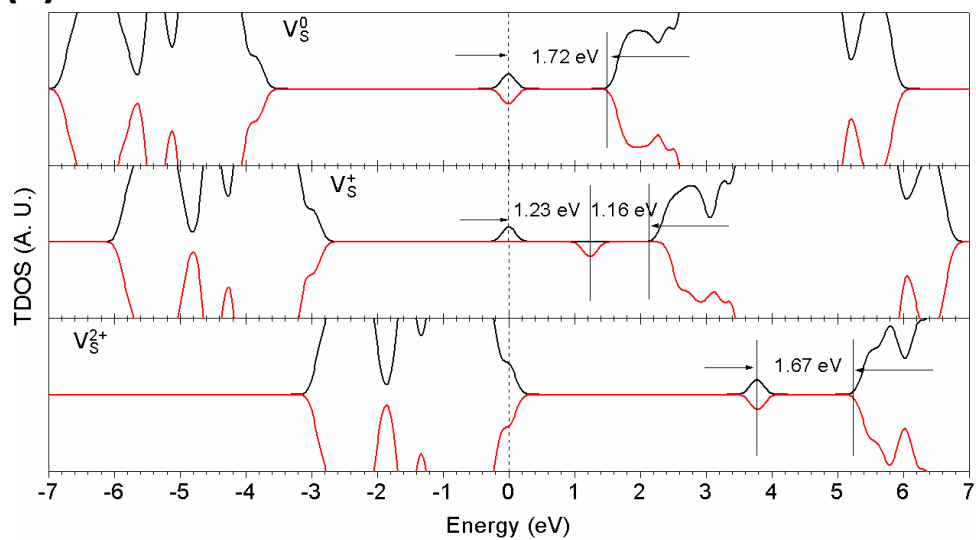

**(b)**

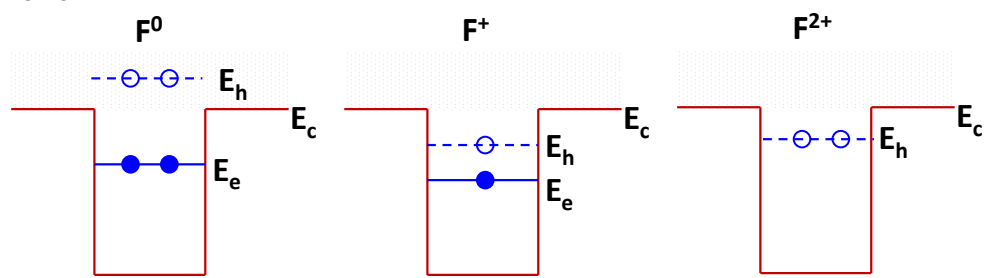

**(c)**

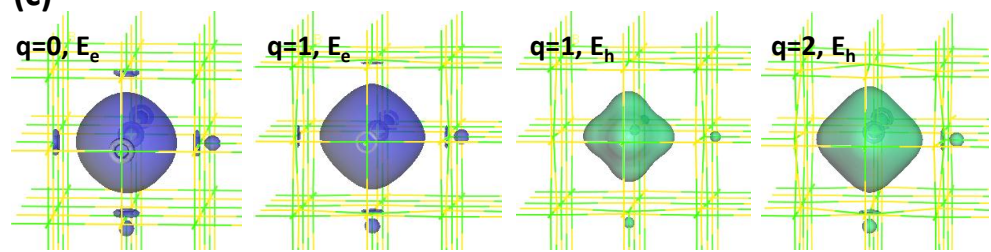

**(d)**

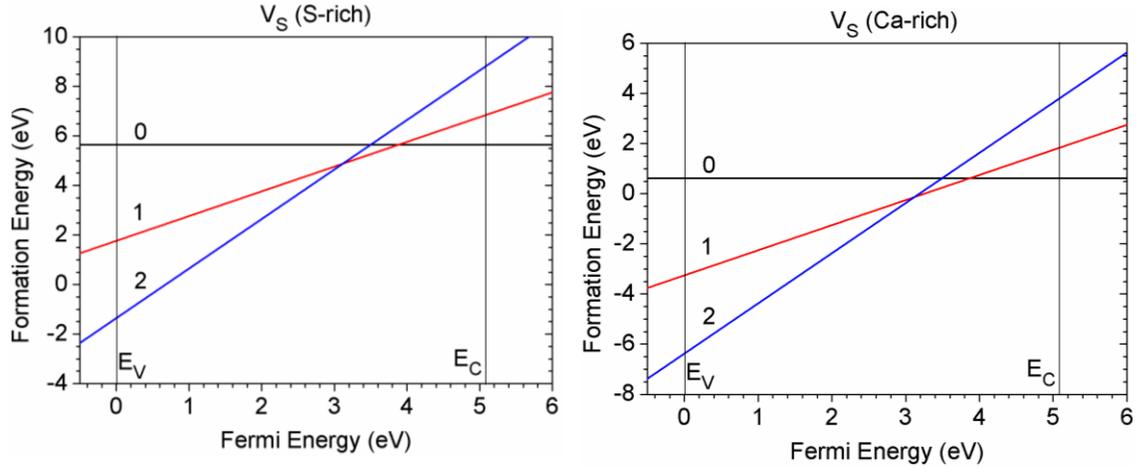

**Figure 2.** (a) TDOSs of $V_S$ in neutral ($V_S^0$), singly positive ($V_S^+$), and doubly positively ($V_S^{2+}$). The dashed line denotes the highest occupied level for electrons. (b) Schematic single particle level of electron and hole levels for describing the F center defects. (c) localized electron and hole orbitals at the relaxed $V_S$ sites (Ca=green, S=yellow). (d) Formation energy of $V_S$ under S and Ca rich chemical potential limits.

**Figure 3.**

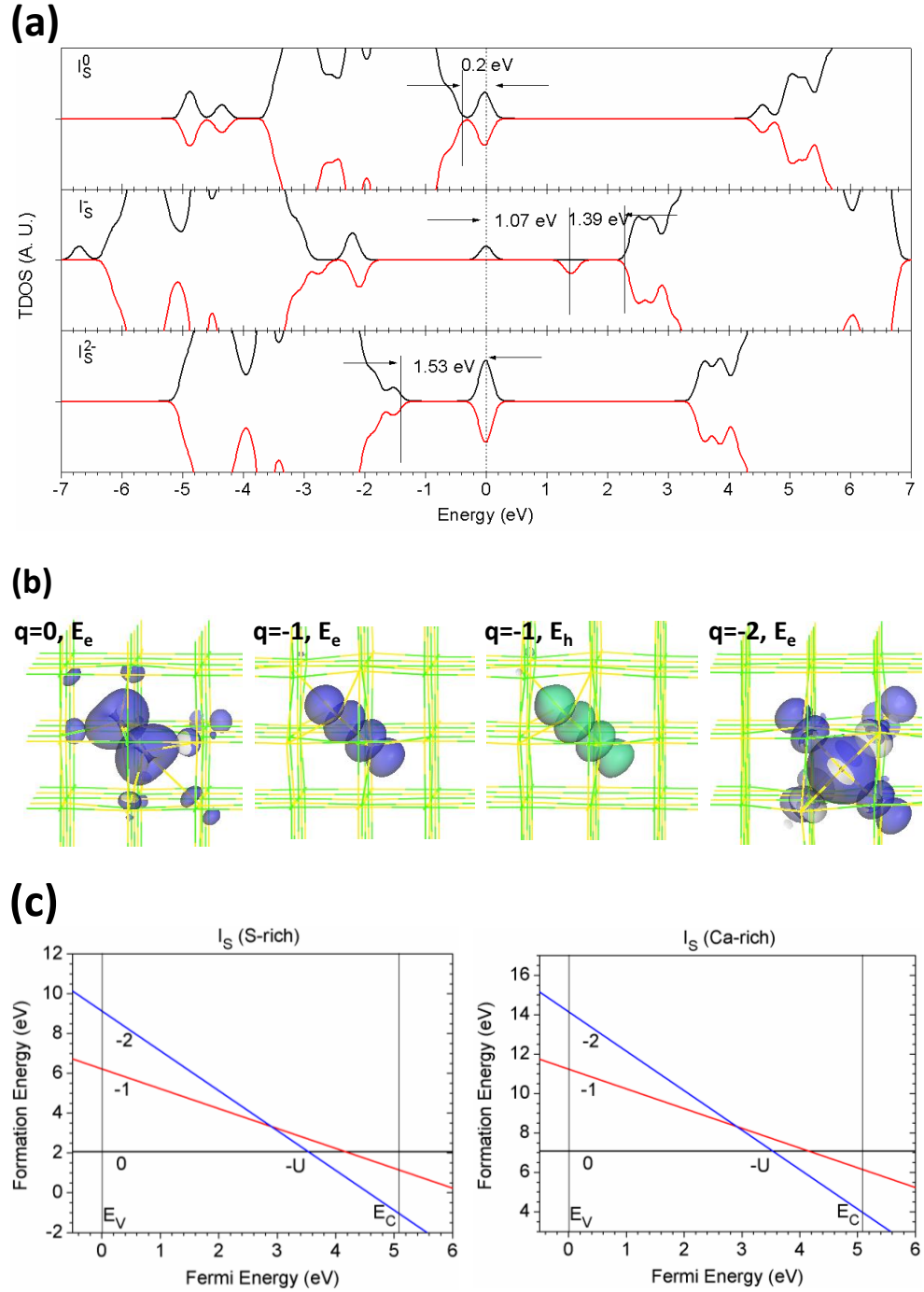

**Figure 3.** (a) TDOSs of $I_S$ in neutral ($I_S^0$), singly negative ($I_S^-$), and doubly negative ($I_S^{2-}$). The dashed line denotes the highest occupied level for electrons. (b) localized electron and hole orbitals at the relaxed $I_S$ sites (Ca=green, S=yellow). (c) Formation energy of $I_S$ under S and Ca rich chemical potential limits. The "–U" denotes the negative-$U_{eff}$ center for $I_S$ at (-2/0) state.

**Figure 4.**

**(a)**

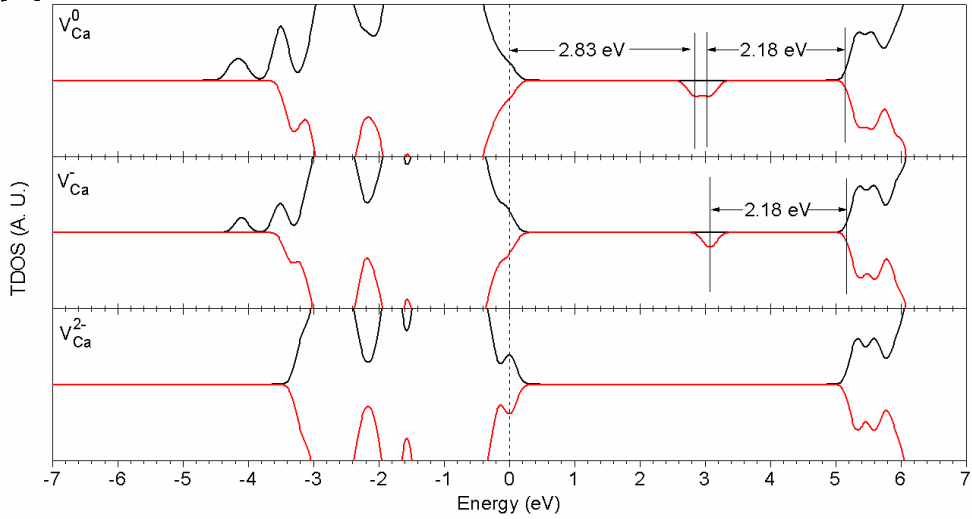

**(b)**

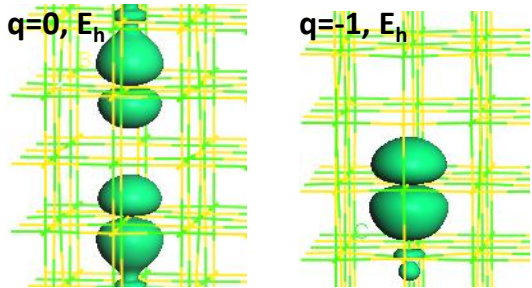

**(c)**

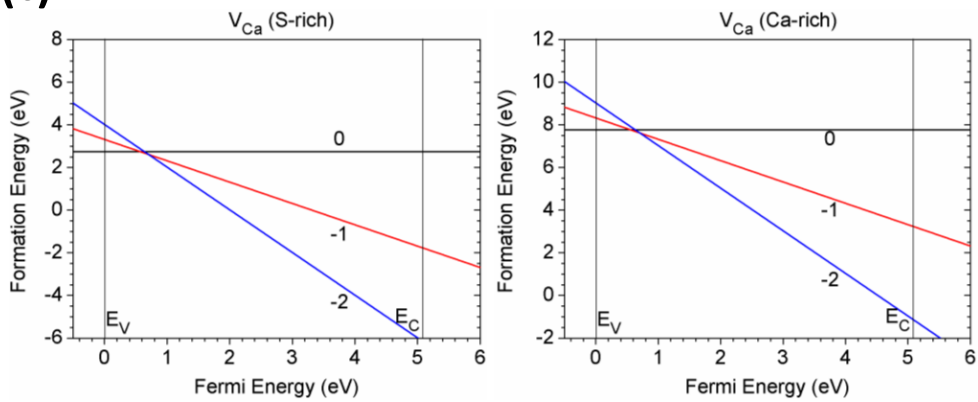

**Figure 4.** (a) TDOSs of $V_{Ca}$ in neutral ($V_{Ca}^0$), singly negative ($V_{Ca}^-$), and doubly negative ($V_{Ca}^{2-}$). The dashed line denotes the highest occupied level for electrons. (b) localized electron and hole orbitals at the relaxed $V_{Ca}$ sites (Ca=green, S=yellow). (c) Formation energy of $V_{Ca}$ under S and Ca rich chemical potential limits, showing a "+$U_{eff}$" defect with a shallow acceptor-trap feature.

**Figure 5.**

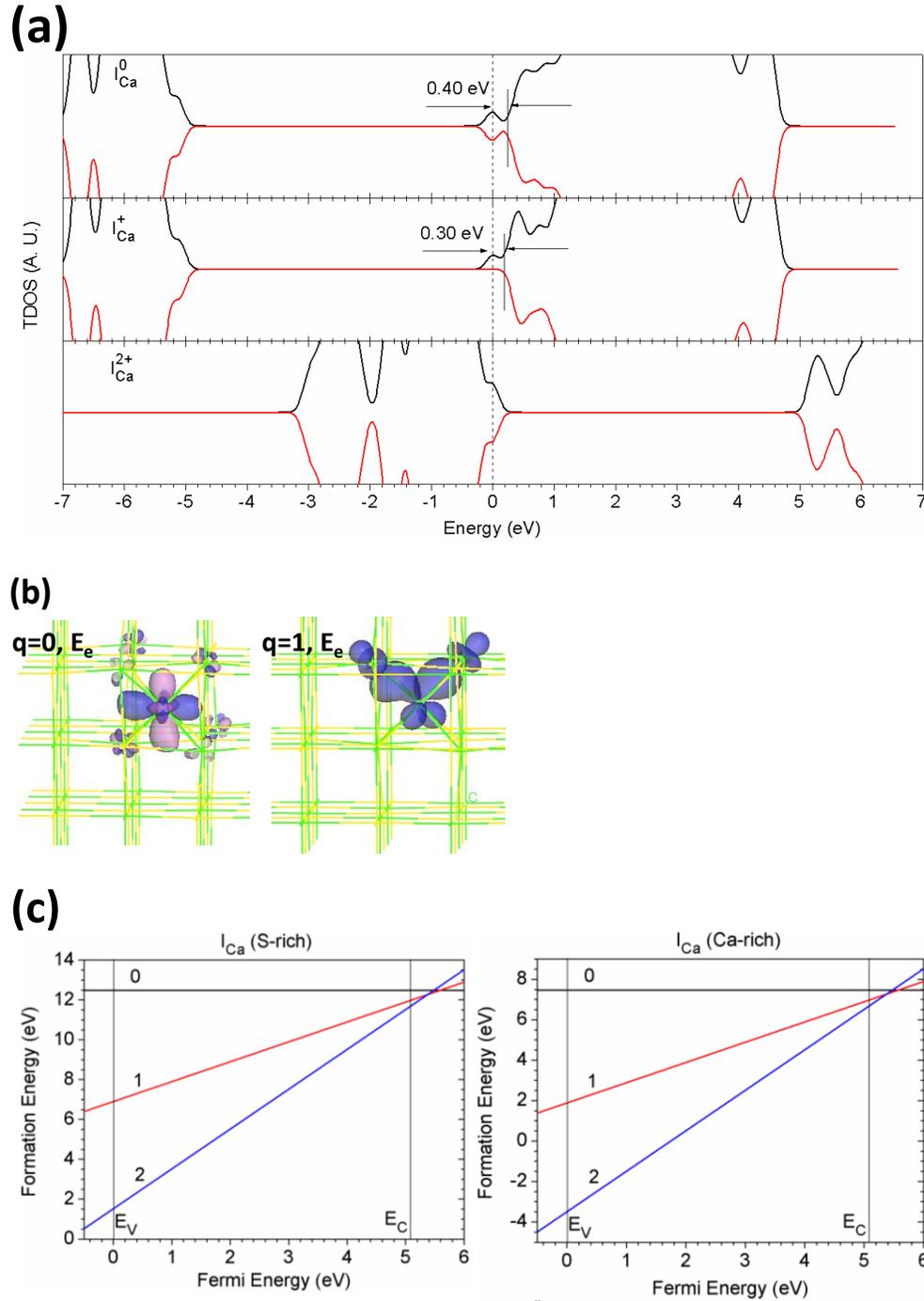

**Figure 5.** (a) TDOSs of $I_{Ca}$ in neutral ($I_{Ca}^0$), singly positive ($I_{Ca}^+$), and doubly positive ($I_{Ca}^{2+}$). The dashed line denotes the highest occupied level for electrons. (b) localized electron orbitals at the relaxed $I_{Ca}$ sites, showing a hybridized d-orbital feature (Ca=green, S=yellow). (c) Formation energy of $I_{Ca}$ under S and Ca rich chemical potential limits, showing a "+$U_{eff}$" defect with a shallow donor-trap features.

**Figure 6.**

**(a)**

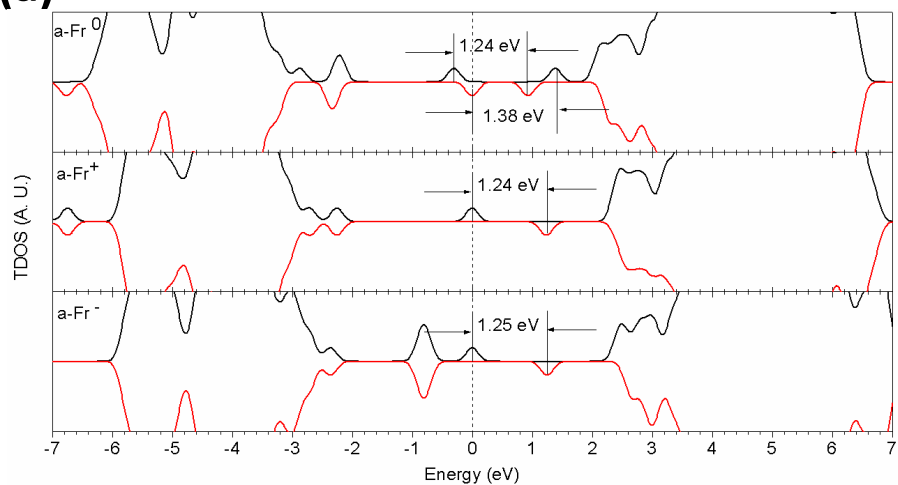

**(b)**

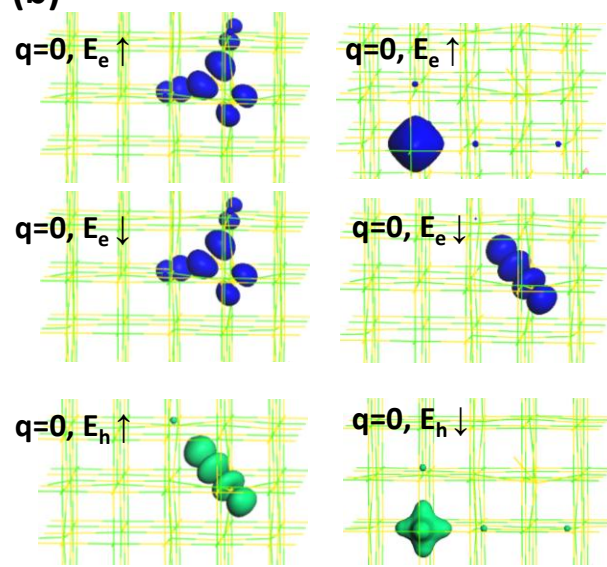

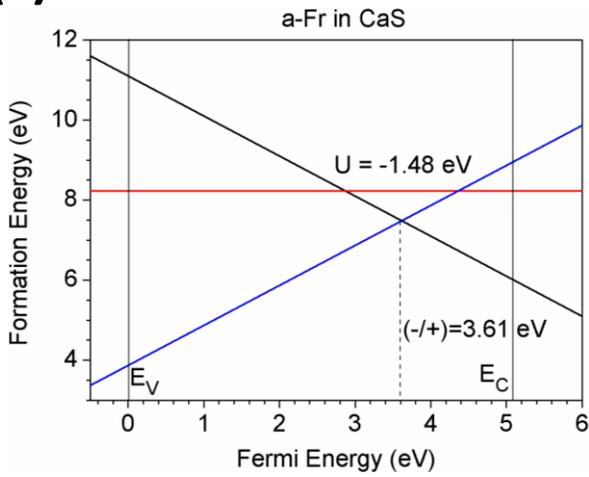

**Figure 6.** (a) TDOSs of a-Fr in neutral (a-Fr$^0$), singly positive (a-Fr$^+$), and singly negative (a-Fr$^-$). The dashed line denotes the highest occupied level for electrons. (b) localized electron orbitals at the relaxed a-Fr$^0$ sites, (Ca=green, S=yellow). (c) Formation energy of a-Fr$^0$ showing a "-$U_{eff}$" defect with transition energy of 3.61 eV for the state (-/+).

**Figure 7.**

**(a)**

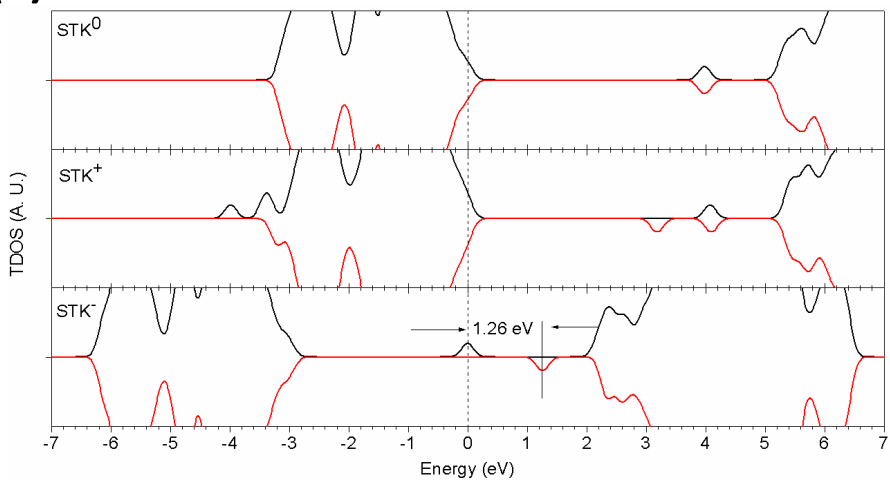

**(b)**

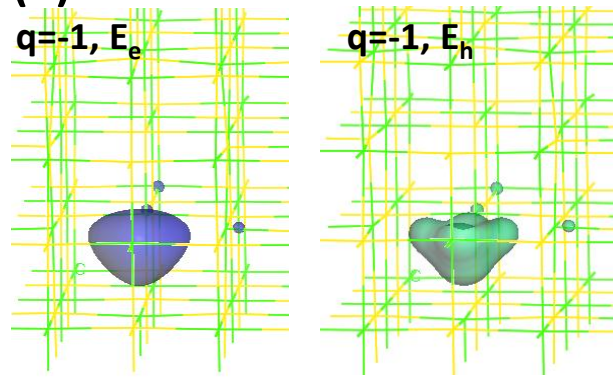

**Figure 7.** (a) TDOSs of Schottky defect pair in neutral (STK$^0$), singly positive (STK$^+$), and singly negative (STK$^-$). The dashed line denotes the highest occupied level for electrons. (b) Localized electron and hole orbitals at the relaxed STK$^-$ sites, (Ca=green, S=yellow).

**Figure 8.**

**(a)**

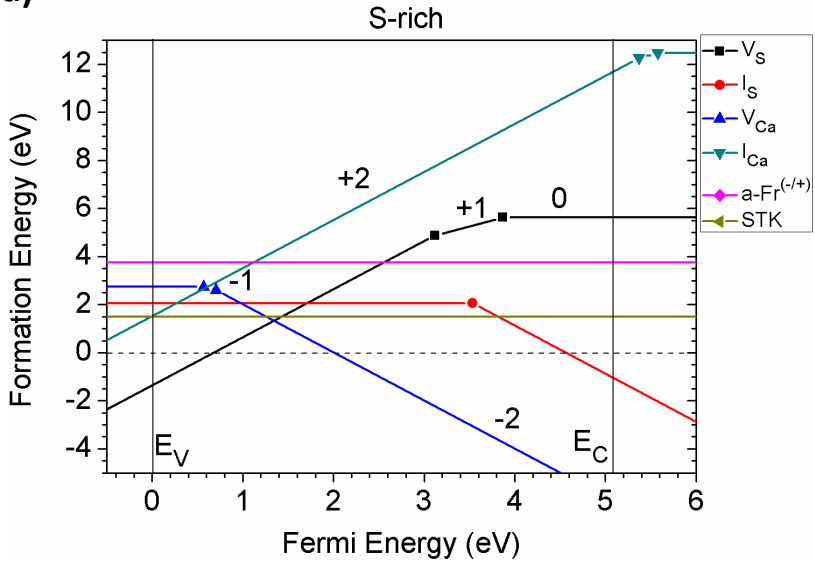

**(b)**

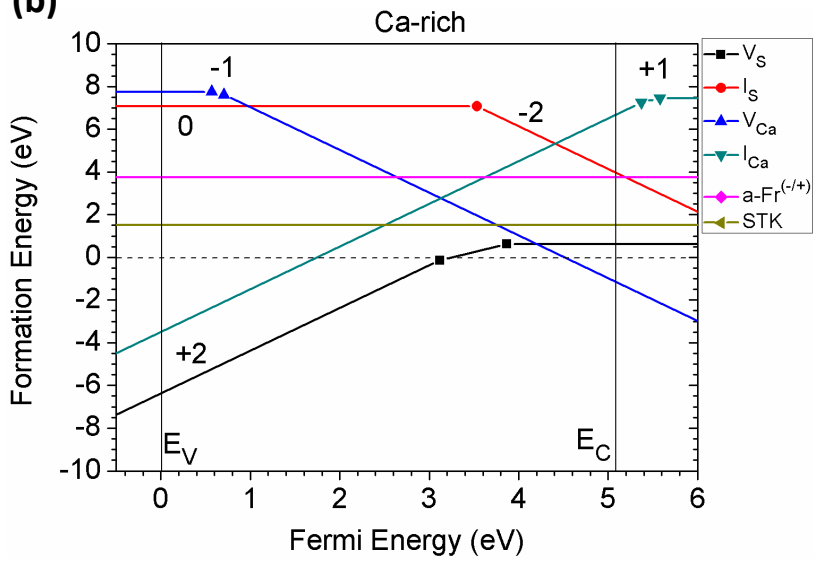

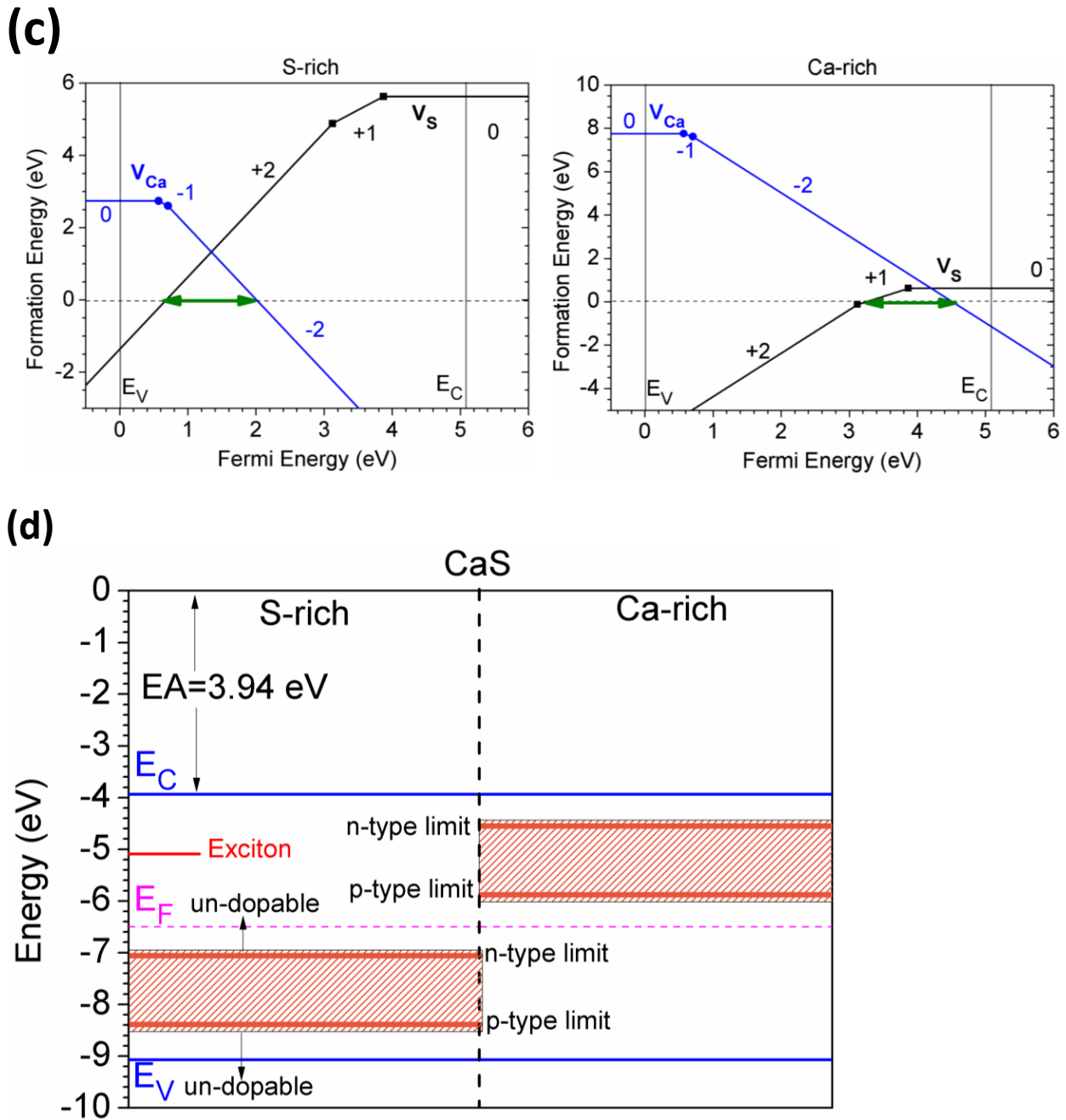

**Figure 8.** Summary of the native point defects in CaS under S-rich (a) and Ca-rich (b) chemical potential limits. (c) The doping limit energy determined by the native point defect formation energies ($V_S$ and $V_{Ca}$ in CaS), the green arrow indicates the dopable range. (d) Schematic band diagram of valence and conduction bands ($E_V$ and $E_C$), ideal Fermi level ($E_F$), doping limit (shaded area), exciton level by native point defects, and electron affinity energy (EA), against the vacuum level (0 eV).